\documentclass{cta-author}

{}
{}
{}

\usepackage{cite} 
\usepackage{graphics} 
\usepackage{subfig} 
\usepackage{amssymb}
\usepackage{makecell}
\usepackage{comment}
\usepackage{tabularx,booktabs,caption,ragged2e}
\newcolumntype{Y}{>{\RaggedRight\arraybackslash}X}
\usepackage{tikz} 
\usetikzlibrary{arrows, shapes, positioning, shadows, trees}

\begin{document}

\title{A systematic review on the role of artificial intelligence in sonographic diagnosis of
thyroid cancer: Past, present and future}

\author{\au{Fatemeh Abdolali$^{1,2}$}
\au{Atefeh Shahroudnejad$^{1,2}$}
\au{Abhilash Rakkunedeth Hareendranathan$^2$}
\au{Jacob L Jaremko$^{1,2}$}
\au{Michelle Noga$^{1}$}
\au{Kumaradevan Punithakumar$^1$}
}

\address{\add{1}{Department of Radiology and Diagnostic Imaging, University of Alberta Hospital, Edmonton, Alberta, Canada, T6G2B7}
\add{2}{MEDO.ai, Dual Headquarters at Singapore and Edmonton, Alberta, Canada,T5J4P6}
\email{abdolali@ualberta.ca}}

\begin{abstract}
Thyroid cancer is common worldwide, with a rapid increase in prevalence across North
America in recent years. While most patients present with palpable nodules through
physical examination, a large number of small and medium-sized nodules are detected
by ultrasound examination. Suspicious nodules are then sent for biopsy through fine
needle aspiration. Since biopsies are invasive and sometimes inconclusive, various research groups have tried to develop computer-aided diagnosis systems. Earlier approaches along these lines relied on clinically relevant features that were manually identified by radiologists. With the recent success of artificial intelligence (AI), various new methods are being developed to identify these features in thyroid ultrasound automatically. In this paper, we present a systematic review of state-of-the-art on AI application in sonographic diagnosis of thyroid cancer. This review follows a methodology-based classification of the different techniques available for thyroid cancer diagnosis. With more than 50 papers included in this review, we reflect on the trends and challenges of the field of sonographic diagnosis of thyroid malignancies and potential of computer-aided diagnosis to increase the impact of ultrasound applications on the future of thyroid cancer diagnosis. Machine learning will continue to play a fundamental role in the development of future thyroid cancer diagnosis frameworks.
\end{abstract}

\maketitle

\section{Introduction}
The incidence of thyroid cancer has increased at an alarming rate in the United States with 52,070 cases diagnosed in 2019, out of which 2170 cases resulted in death [1]. It is the most common cancer in American women aged between 20 and 34 [1]. The corresponding numbers in Canada are equally significant with the Canadian Cancer Statistics estimating that 8,200 Canadians (6100 women and 2100 men) would be diagnosed with thyroid cancer resulting in 230 deaths in 2019 [2]. The incidence of thyroid cancer has steadily increased from 1970, with the most substantial increase in middle-aged women [3]. \par
Ultrasound (US) is a fast, safe and inexpensive imaging technique that can provide a complete visualization of thyroid nodules. Imaging is generally performed in both transverse and sagittal orientation using ultrasound probes in the 7 - 15 MHz range. Thyroid nodules have several distinguishable characteristics on ultrasound such as shape, size, echogenicity (brightness) and echotexture (composition). For example, features such as microcalcification or taller-than-wide nodule shape can be predictors of malignancy, whereas spongiform appearance can be representative of benign case [4]. \par
The critical challenge in image aided thyroid nodule diagnosis and reporting of thyroid nodules is the extraction of optimal sets of features from ultrasound images that differentiate malignant from benign nodules. In order to standardize the reporting of thyroid nodules, American College of Radiology introduced a Thyroid Imaging, Reporting and Data System (TIRADS) [5]. The goal of TIRADS is to provide standardized recommendations for the management of thyroid nodules on the basis of well-defined clinical features for every lesion. TIRADS identifies six categories for a given nodule: composition, echogenicity, shape, size, margins and echogenic foci. Based on the scores of each of these features, the nodule is classified as benign, minimally suspicious, moderately suspicious or highly suspicious of malignancy.Various studies have reported increased agreement among users while using TIRADS. A parallel approach aimed at reducing the variability in thyroid nodule reporting is based on computer-aided diagnosis (CAD) systems that usually categorize nodules as malignant or benign. These systems are mostly trained on retrospectively collected biopsy data and generally aim to reduce false positives that would otherwise result in unnecessary biopsies and over-diagnosis [6]. \par
In this paper, we review various artificial intelligence (AI)-based CAD systems that have been proposed for thyroid ultrasound nodule analysis, including segmentation, detection and classification. Earlier reviews have focused on specific methods such as feature-based [7], deep learning [8], linear [9] and non-linear [10] approaches (see Table 1). We aim to provide a broader perspective in our discussion and compare the above-mentioned approaches to one another. Our review also includes more recent publications in each of these approaches. This paper presents, for the first time, a systematic review of the existing literature on sonographic diagnosis of thyroid nodules, which covers both classical and deep learning methods. \par
Figure 1 shows the histogram of the papers introducing novel machine learning-based techniques for thyroid nodule diagnosis. Inclusion and exclusion criteria are summarized in Table 2. With more than 50 articles included in this survey, we focus on the methodological aspects of those works. We propose a structured analysis of the different approaches used for ultrasound-based thyroid tissue quantification. The paper is organized as follows. Section 2 presents an overview of the parameterization models for thyroid ultrasound nodule analysis. In Section 3, we will discuss the main findings of previous works, limitations, challenges and future trends that are specific to thyroid US images. Conclusions are presented in Section 4.

\begin{figure*}[htbp]
    \centerline{\includegraphics[width=1.8\columnwidth,trim=0.2cm 0 0.3cm 0, clip]{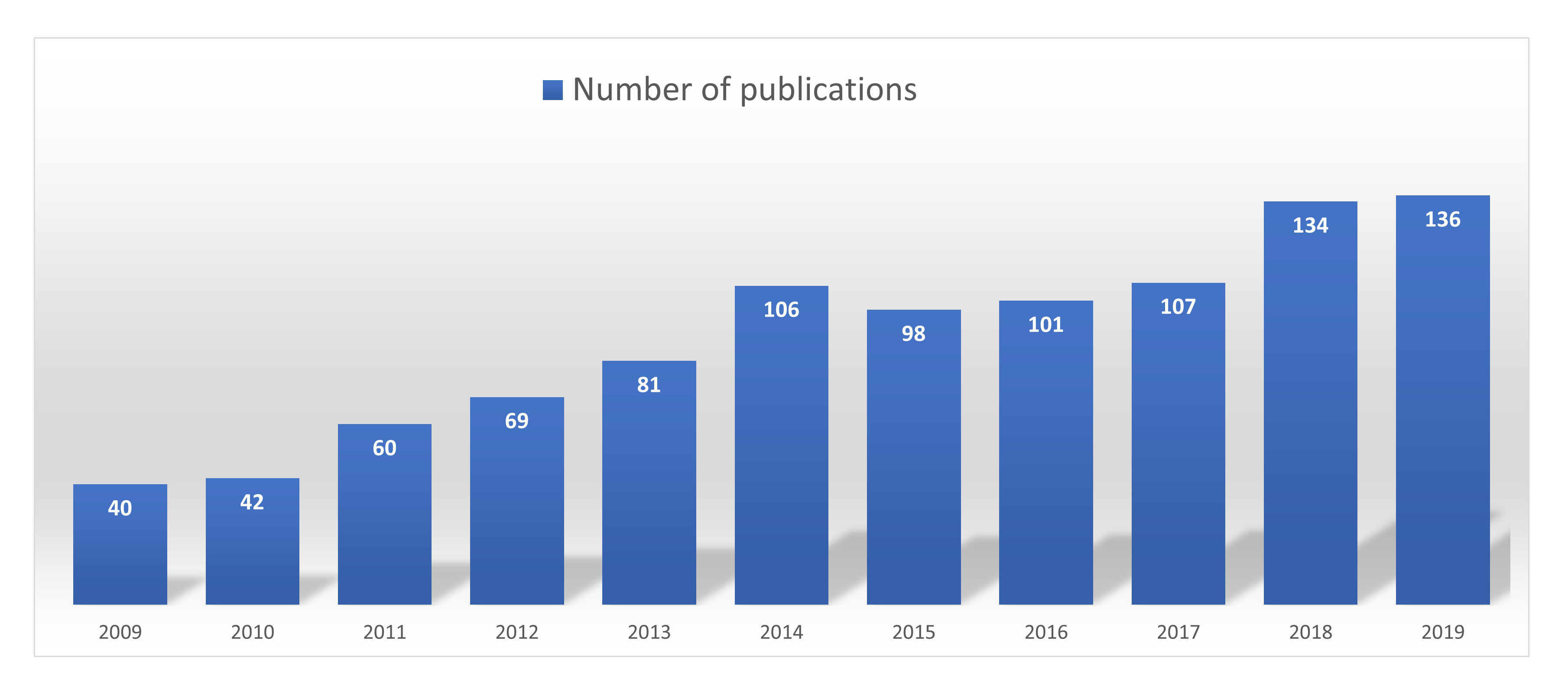}}
    \caption{Publications on sonographic diagnosis of thyroid per year based on PubMed and Scopus search engines.}
    \label{fig:histFinal}
\end{figure*}

\begin{table}[htbp]
	\caption{Overview of review articles in diagnosis of thyroid cancer}
	\setlength{\tabcolsep}{3pt}
	\label{tab:table1}
	{
	\begin{tabular}{ l c c c c c }
		\hline
		\textbf{Article} & \textbf{Year} & \textbf{Study purpose} & \textbf{\makecell{Imaging \\ modalities}} & \textbf{\makecell{Number \\ of papers}} \\
		\hline
		Acharya \emph{et al.}[7]& 2014 & \makecell{Summarizing studies \\ on malignancy \\detection} & US & 57 \\
		\hline
		Khachnaoui \emph{et al.}[8] & 2018 & \makecell{Summarizing deep\\ learning-based \\studies on \\thyroid diagnosis} & US & 8\\
		\hline
		Sollini \emph{et al.}[9] & 2018 & \makecell{Summarizing studies \\ on texture analysis \\with and without \\ CAD}& \makecell{US \\ CT \\ MRI \\ PET} & 66\\
		\hline
		Verburg \emph{et al.}[10] & 2019 & \makecell{Meta-analysis study \\for US classification\\ of thyroid cancer} & US & 10 \\
		\hline
	\end{tabular}}
	\label{tbl:overview}
\end{table}	

\section{Review of methodological approaches for thyroid ultrasound analysis}
The aim of this section is to provide a structured reference guide for the different techniques of thyroid ultrasound analysis from a methodological point of view. All these works share a common goal: use CAD for thyroid parameterization. Figure 2 shows the general block diagram of a CAD system for sonographic diagnosis of the thyroid cancer. At first, some preprocessing steps such as rescaling pixel intensities and noise reduction are performed and region-of-interest (ROI) is selected from every US image. Then, feature extraction approach is applied. Finally, thyroid gland or nodule region is segmented so that the prediction about the malignancy could be performed. To find relevant literature sources, PubMed and Scopus search engines were utilized using the query string: (a) “artificial intelligence” or “deep learning” or “segmentation” or “detection” and (b) “thyroid”. The last search was conducted on 12 December 2019. The search produced 997 results from search queries. Titles and abstracts were reviewed to determine whether they were suitable for this review. We adopted a set of inclusion and exclusion criteria (see Table 2). The articles were downloaded, and the titles, abstract and methods sections were reviewed in order to verify that the articles met the aforementioned criteria. A total of 52 articles met the criteria for inclusion in the final review. Overall, our review was aimed at describing recent artificial intelligence achievements in sonographic diagnosis of thyroid nodules. \par
Based on the papers surveyed, we identified 2 major categories: 1) clinical features-based models; 2) machine learning-based models; and several other sub-categories of methods.

\begin{table}[b]
    \centering
	\caption{Inclusion and exclusion criteria}
	\label{table2}
	\begin{tabularx}{8cm}{@{} lY @{}}
	\toprule
	\textbf{Inclusion criteria} & \textbf{Exclusion criteria} \\
	\midrule
	(a) Published between 2009 and 2019. & (a) Studies that utilized imaging modalities other than US.\\
    (b) Peer-reviewed article. & (b) Studies that focused only on case reports.\\
	\bottomrule
	\end{tabularx}
\end{table}

\tikzstyle{decision} = [diamond, draw, fill=blue!20, 
    text width=4.5em, text badly centered, node distance=3cm, inner sep=0pt]
\tikzstyle{block} = [rectangle, draw, fill=lightgray, 
    text width=6em, text centered, rounded corners, minimum height=4em]
\tikzstyle{line} = [draw, -latex']
\tikzstyle{cloud} = [draw, ellipse,fill=red!20, node distance=3cm,
    minimum height=2em]

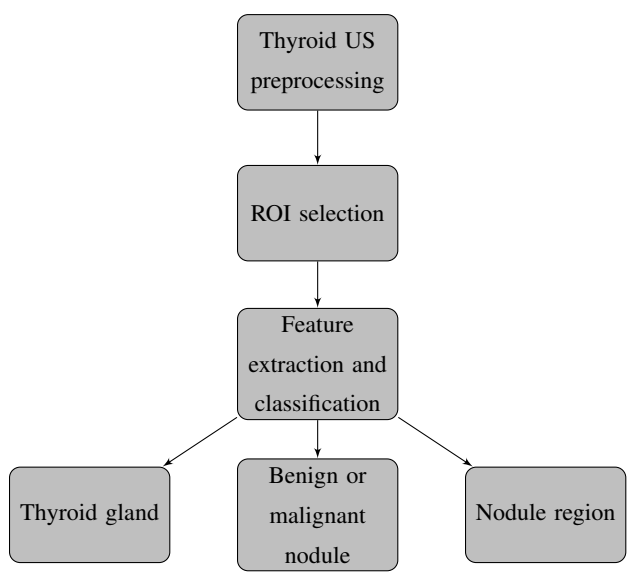
\begin{figure}[b]
\begin{tikzpicture}[node distance = 2cm, auto]
    \node [block] (init) {Thyroid US preprocessing};
    \node [block, below of=init] (identify) {ROI selection};
    \node [block, below of=identify] (evaluate) {Feature extraction and classification};
    \node [block, below of=evaluate] (Benignnodule) {Benign or malignant nodule};
    \node [block, left of=Benignnodule, node distance=3cm] (Thyroid) {Thyroid gland};
    \node [block, right of=Benignnodule, node distance=3cm] (region) {Nodule region};
    \path [line] (init) -- (identify);
    \path [line] (identify) -- (evaluate);
    \path [line] (evaluate) -- (Benignnodule);
    \path [line] (evaluate) -- (region);
    \path [line] (evaluate) -- (Thyroid);
\end{tikzpicture}
	\caption{\footnotesize Block diagram of a typical CAD system for sonographic diagnosis of thyroid cancer}
	\label{fig:BlockDiagram}
\end{figure}
	
\subsection{Clinical features-based Models}
Figure 3 shows an example of an image of the thyroid gland obtained by ultrasound. As can be appreciated, the correct interpretation of ultrasound images requires an expert radiologist. Zahang et al. [11] performed malignancy detection based on both conventional US and real-time elastography (RTE). A set of clinical features, i.e., echogenicity, margins, internal composition, aspect ratio, vascularity, hypoechoic halo, calcifications and real-time elastography grade were fed to nine different conventional classifiers. It was concluded that random field classifier based on US and RTE features outperformed other classifiers. In [12], ReliefF feature selection was utilized to select the most discriminative sonographic features. Selected features were fed into extreme learning machine (ELM), support vector machine (SVM) and neural network. In another research [13], sonographic features were fed into statistical classifiers to differentiate malignant thyroid nodules.

\begin{figure}[t]
    \centerline{\includegraphics[width=\columnwidth,trim=0.4cm 0.5cm 0.5cm 0.9cm, clip]{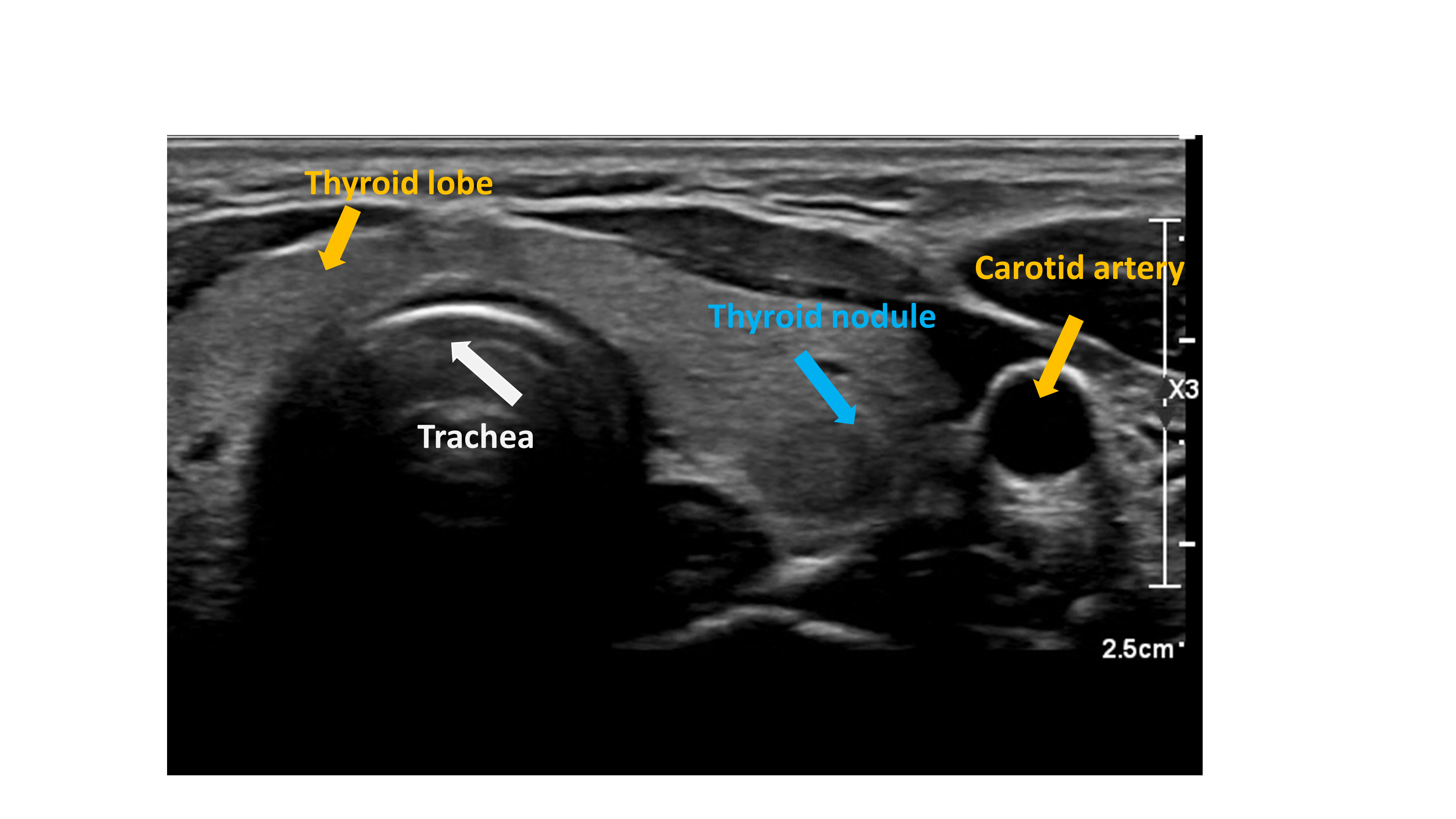}}
    \caption{Transverse US image showing a thyroid nodule in the right thyroid lobe and adjacent anatomical structures.}
    \label{fig:ThyroidData}
\end{figure}

\subsection{Machine learning-based models}
Several promising machine learning-based models were proposed for sonographic diagnosis of thyroid cancer. Based on the papers surveyed, we identified 2 major categories: 1) classical features-based models; 2) Deep learning models, as discussed below.

\subsubsection{Classical features-based models}
The use of classical features has been extensively studied in a variety of applications, including automatic segmentation of thyroid and nodules, thyroid cancer detection and classification. Their ability to integrate knowledge about location, size and shape of the thyroid organ or nodule has led to some of the promising segmentation [14–18] and detection methods [12, 19, 20]. Early classical features-based frameworks used discriminative statistical and texture features, i.e., wavelet transformation [21, 22], histogram feature [18], block difference of inverse probabilities (BDIP) and normalized multiscale intensity difference (NMISD) features [18] and co-occurrence matrix [21]. Then, feature selection techniques, i.e., minimum redundancy maximum relevance (mRMR) [23] and T-test [24] were utilized to select those features which contribute most to the model. The selected feature set was fed to classifier, i.e., SVM [12, 19, 20, 23–26], Random forest [11, 15, 19] , neural network [11, 12, 22] or Adaboost [11, 21]. These approaches can be divided into thyroid or nodule segmentation and thyroid cancer detection, as discussed below.

\paragraph{Thyroid and nodule segmentation}
Several methods for automated segmentation of thyroid or nodules were proposed in the literature. Illanes et al. utilized a combination of wavelet texture extraction and parametrical modelling for US thyroid segmentation [14]. In [15], an algorithm based on iterative random walks solver and gradient vector flow based interframe belief propagation technique was developed to segment thyroid, lumen and external elastic laminae. Koundal et al. [16] proposed Spatial Neutrosophic Distance Regularized Level Set method based on Neutrosophic L-Means clustering to segment thyroid nodules. Later, they used Neutrosophic Nakagami Total Variation and small rectangular region extraction as pre-processing steps [27] to improve the segmentation results. Multi-organ segmentation of thyroid, carotid artery, muscles and trachea was performed by Narayan et al. [17] using speckle patch similarity estimation. In [18], Radial Basis Function (RBF) neural network was trained to classify each block of US image into thyroid gland and non-thyroid gland pixels.

\paragraph{Thyroid cancer detection and classification}
The use of classical features has been extensively studied for thyroid cancer detection. Two threshold binary decomposition was utilized in [19] for patch based feature extraction. The extracted features were fed into random forest and SVM classifiers. Yu et al. [20] developed a framework based on region-based active contours and texture features extraction. The best performance was achieved by combining Artificial neural networks (ANN) and SVM. In other work, Gabor features were extracted from high resolution US images [24]. Locality sensitive discriminant analysis was utilized for feature space reduction. The selected features were fed into SVM, k-nearest neighbors (KNN) and multilayer perceptron (MLP) classifiers. Wavelet transformation was utilized as feature extraction method in several studies [21, 22]. In another study, different linear and nonlinear machine learning algorithms were evaluated for classification of thyroid nodules [28].

\subsubsection{Deep learning models}
Deep neural networks are representation learning methods which extract features from input data and use these features to perform machine learning tasks such as segmentation or detection. These approaches can be divided into thyroid or nodule segmentation and thyroid cancer detection, as discussed below.

\paragraph{Thyroid and nodule segmentation}
Poudel et al. [29] compared the performance of four popular segmentation methods, i.e., active contour without edges, graph cut, pixel based classifier and random forest classifier with 3D U-net [30] on Opencas thyroid dataset [31]. They concluded that 3D U-net outperforms other methods in terms of accuracy. Several deep learning architectures were utilized for nodule segmentation in the literature. In [32], mark-guided U-net [33] was proposed for segmentation of thyroid nodules. Four landmark points corresponding to the major and minor axes of a nodule were determined manually to guide the training and inference of U-Net. A deep CNN framework with multiple intermediate layers was introduced in [34] for thyroid nodule segmentation. The proposed CNN contains 15 convolutional layers and two pooling layers. The main drawback of the proposed approach is that some thyroid nodules with complicated back-ground cannot be accurately segmented. An 8 layer fully convolutional network was proposed in [35] for thyroid nodule segmentation. The proposed framework outperformed U-net on the local test dataset.

\paragraph{Thyroid cancer detection and classification}
Several deep learning based frameworks were proposed by different groups of scientists in the recent years. A research group
evaluated the accuracy of thyroid cancer diagnosis using deep learning [10]. Authors utilized the dataset from three different Chinese
hospitals and concluded that deep learning approach resulted in slightly inferior sensitivity and substantially improved specificity. A deep learning approach inspired by faster R-CNN [36] was proposed in [37] for thyroid papillary cancer detection. Using layer concatenation strategy, more detailed features of low resolution US images were extracted. An end-to-end detection network based on integrating YOLOv2 [38] and Resnet v2-50 [39] was employed in [40]. A retrospective, multicohort, diagnostic study using ultrasound images sets from three hospitals in China was performed in [41]. The CNN model was developed by ensembling Resnet50 [42] and Darknet19 [43] models. Song et al. [44] proposed a deep learning framework based on Inception-V3 network model [45] for diagnosis of thyroid nodules. The model was trained on a small local dataset which consists of cropped nodules by a clinician. Two pre-trained CNN models, i.e., “imagenet-vgg-verydeep-16” and “imagenet-vgg-f”, were used in thyroid nodule malignancy detection [46]. To train the CNN with their local data, the ROIs were extracted by a radiologist from each US image.

\section{Discussion}
Our review of thyroid diagnosis techniques from US images has shown that accurate interpretation of ultrasound images is based on clinical feature-based models. However, recent advances in artificial intelligence have introduced efficient and flexible machine learning techniques applied to the analysis of thyroid US images. Machine learning will continue to play a fundamental role in the development of future thyroid cancer diagnosis frameworks. In this section, we discuss the applications of artificial intelligence in sonographic diagnosis of thyroid cancer, as well as the limitations and opportunities in this growing field.

\subsection{Analysis by reported results and main findings}
Several CAD systems for sonographic diagnosis of thyroid cancer have been developed, from clinical feature-based systems to machine learning-based systems. See Table 5 and Table 3 for a detailed list of the publications reviewed in this paper. In this section, we compare the previous works based on the reported results. For nodule segmentation,  several methods [16, 32, 35, 47] achieved Dice scores higher than 90\%. The best performance was achieved using Spatial Neutrosophic Distance Regularized level set method based on Neutrosophic L-Means clustering [16]. For thyroid segmentation, several methods [14, 17, 18, 29] achieved Dices score higher than 85\%. Chang et al. [18] proposed a framework based on Haar Wavelet features and RBF neural network to classify each block of images into thyroid gland and non-thyroid gland. They achieved mean Dice score of 96.52\% for thyroid segmentation. Several groups utilized deep learning or classical features for thyroid nodule classification. In a recent study, spatial domain features based on deep learning and frequency domain features based on Fast Fourier transform were combined for classifying the input thyroid images into either benign or malign cases [48]. The best classification accuracy was reported using multi-task cascade VGG-16 model [49]. \par
The main limitations of previous works are discussed below.

\subsection{ Limitations, Challenges and Future Trends}
During the last 2 decades, sonographic imaging of the thyroid gland was the most valued imaging method. However, there are several cases where the radiologist cannot distinguish between benign and malignant nodules with complete certainty. The reliability of diagnosis depends on the quality of US images and the expertise of the medical experts who interpret US images. Machine learning methods can help to extract features beyond human perception which leads to more effective characterizationof thyroid anatomy. \par
Previous works focused on techniques for analysis of one nodule per US image. If one or more nodules are detected within the thyroid gland, it is not clear whether these techniques can be accurate enough. More experiments should be performed to evaluate the accuracy of previous techniques in such scenarios. \par
The limited availability of annotated ultrasound data has been a problem in automated sonographic diagnosis of thyroid cancer.This limitation becomes particularly relevant when implementing CAD systems for thyroid nodule classification, where large datasets are needed to characterize location and texture of thyroid gland. The need for large datasets is essential for the development and validation of new CAD systems for thyroid. Moreover, this represents a major obstacle to realize the full potential of deep learning based techniques. Although publicly available datasets with manual annotations of thyroid exist, the number of cases is limited to a couple of hundred cases at best. Public data repositories, i.e., DDTI [31] and Opencas dataset [50], provide open access to detailed manually-guided expert annotations of thyroid structure and nodules. Collection of a large comprehensive dataset is critical to develop future CAD systems that are robust to pathology.

\section{Conclusions}
In this paper, we presented the first systematic review of CAD systems for sonographic diagnosis of thyroid cancer. The continuing progress of machine learning have favored the development of complex and comprehensive models for sonographic diagnosis of thyroid nodules. As shown in this survey, the automatic parameterization of thyroid has been approached from different perspectives, (e.g. detection, segmentation and classification) using various methodologies. The categorization of approaches in this paper provides a reference guide to the current techniques available for the analysis of thyroid US images. We have also indicated current challenges and future opportunities in CAD systems for analysis of thyroid US images. New efficient machine learning models should embed the anatomical context inherent to the thyroid gland to provide the essential  accuracy for clinical practice.

\section*{Acknowledgment}

This work was supported by Mitacs through the Mitacs Accelerate Program (IT12871).


%


\begin{table*}[b]
	\caption{Overview of applications of deep learning based methods in sonographic diagnosis of thyroid cancer}
	\centering
	\label{table3}
	\resizebox{\textwidth}{!}{\begin{tabular}{| c | c |c | c | c|c|c|c|c|}
		\hline
		\textbf{Citation} & \textbf{Study Purpose} & \multicolumn{3}{c|}{\textbf{Dataset}} &\textbf{ Method} & \textbf{Measures }& \textbf{Results or main findings}&\textbf{TI-RADS}\\
		& &\textbf{General information} & \textbf{Number of patients} & \textbf{2D or 3D} & & & & \\
		\hline
		[40]-2019 & \makecell{Nodule detection \\ and classification} & \makecell{Local \\ training (2450 benign \\ cases- 2557 malignant cases)\\test (351 with nodules-\\ 213 normal cases)} & 276 for test & 2D & \makecell{Integration of Resnetv2-50 and YOLOv2}& \makecell{ROC\\Sensitivity\\PPV\\NPV\\Accuracy\\Specificity}& \makecell{ROC=90.5\%\\Sens=95.22\%\\PPV=80.99\%\\NPV=90.31\%\\Acc=90.31\%\\Spec=89.91\%}&  \checkmark\\
		\hline
		[41]-2019 & Thyroid cancer diagnosis & \makecell{training(312399)\\validation(20386)} & \makecell{training(42952)\\validation(2692)} & 2D & \makecell{Ensemble DCNN (Resnet50 and Darknet19)}& \makecell{AUC\\Sensitivity\\Specificity}& \makecell{AUC= 92.16\%\\Sens= 87.46\% \\Spec= 86.93\%}&  \checkmark \\
		\hline
		[44]-2019 & \makecell{Thyroid nodule classification} & \makecell{Local (training data: 1358, \\ 670 benign, 688 malignant \\+ 55 internal Korean and \\ 100 external Japanese test sets)} & Not mentioned & 2D & \makecell{Inception-v3 model} & \makecell{Sensitivity\\Negative predictive value} & \makecell{1. Internal set: sensitivity=95.2\%, NPV=95.5\% \\ 2.External set:sensitivity=94.0\%, NPV=90.3\%} &  \checkmark\\
		\hline

		[37]-2018 & \makecell{Cancer regions detection in \\ thyroid papillary carcinoma images} & Local (4670) & 300 & 2D & \makecell{Improved Faster R-CNN} & \makecell{TPR, TNR \\ AUC-ROC} & \makecell{TPR=0.935 \\ TNR=0.815 \\ AUC-ROC=0.938} & \\
		\hline	
			
		[35]-2018 & Nodules segmentation & \makecell{Training data: 200 \\ Test data: 100}& Not mentioned & 2D & \makecell{8-layer fully-convolutional net\\(replace fc with convolution in VGG19)\\+3-deconvolution layers}& IoU&  \makecell{Acc:91\%} &\\
		\hline
		[29]-2018 & Thyroid segmentation &\makecell{1416 images from \\Open-CAS dataset} & 10 & 2D & \makecell{Active Contour without Edges, Graph Cut, \\Pixel-Based Classifier, Random Forest Classifier \\and Convolutional Neural Network (3D U-net)} & \makecell{Dice’s coefficient\\ Hausdorff distance} & \makecell{U-net outperformed other methods in \\ terms of accuracy, Dice=0.876 and \\ Hausdorff distance=7.0 mm} &\\
		\hline
		[49]-2018 & \makecell{Nodule detection }& Local (6228)& 1580 & 2D & \makecell{VGG-16 for extracting feature maps- classification using spatial pyramid module}& AUC-ROC& Acc: 98\% &\\
		\hline
		[32]-2018 & Nodule segmentation & Local (893) & Not mentioned & 2D & \makecell{UNet guided by 4 points on the ROI}& \makecell{Overlap- Dice \\ Accuracy\\ TPF}& \makecell{Dice=94\% \\Acc= 97\% \\ TPF=94\%} & \\
		\hline
		[51]-2018 & Lymph node metastasis detection & \makecell{train (263 benign-286 metastatic)\\ valid (30 benign-33 metastatic)\\ test(100 benign-100 metastatic)} & 604 & 2D & \makecell{Fusing VGG and class activation \\map in one model}& \makecell{Sensitivity\\PPV\\NPV\\Accuracy\\Specificity}& \makecell{Sens=89\% \\PPV=79.5\% \\NPV=87.6\%\\Acc=83\%\\Spec=77\%} & \\
		\hline
		[52]-2018 & \makecell{Nodules classification} & \makecell{1.Local(800 thyroid nodule images)\\2.194 images from DDTI dataset} & \makecell{1.Not mentioned\\2.Not mentioned} & \makecell{1.2D\\2.2D} & \makecell{Nodule ROI detection with \\ VGG-16; \\Benign/Malignant classification \\by training CNN with EM algorithm} & \makecell{Accuracy\\Sensitivity\\Specificity}& \makecell{1.Accuracy=0.88,sensitivity=0.90,\\specificity=0.86\\2.Accuracy=0.80,sensitivity=0.81,\\specificity=0.80} & \\		
		\hline
		[53]-2017 & Nodule classification & DDTI dataset & 298 & 2D & \makecell{Data augmentation +transfer\\ learning using pre-trained ResNet}& \makecell{Accuracy\\Sensitivity\\Specificity}& \makecell{Accuracy=93.75\%\\Sensitivity=93.96\%\\Specificity=92.68\%}& \\		
		\hline
		[54]-2017 & Nodule classification & \makecell{1.Local (164 images) \\ 2.DDTI dataset (428 images)} & \makecell{1. Not mentioned \\ 2. 400} & 2D & \makecell{Fine-tuned GoogLeNet model for feature extraction\\ Random Forest classifier}&  \makecell{classification accuracy, sensitivity, \\specificity and ROC} & \makecell{1. Acc:96.34\%, sensitivity:86\% \\and specificity:99\% \\2.Acc:98.29\%, sensitivity:99.10\% \\and specificity:93.90\%} & \checkmark \\
		\hline
		[34]-2017 & Nodule detection & Local (21523) & 5842 & 2D & \makecell{Nodule segmentation \\ using CNN15 and \\ classification using CNN4}& \makecell{ROC- AUC}& \makecell{AUC = 98.51\%}&  \\
		\hline
		[47]-2017 & Nodule segmentation & Local (22123) & 6242 & 2D & \makecell{Nodule segmentation \\ using CNN15}& \makecell{TP-FP\\Overlap-Dice\\ Hausdorff\\distance}& \makecell{TP=91\%\\FP=6\%\\Over=86\% \\ Dice=92\% \\HD= 62\%}&  \\
		\hline		
		[55]-2017 & Thyroid cancer detection & \makecell{Local(1037 thyroid \\nodule images)} & Not mentioned & 2D & \makecell{Combination of VGG features,\\Histogram of Oriented\\Gradient (HOG) and \\Local Binary Patterns (LBP) \\for classification} & \makecell{Accuracy\\ Sensitivity\\ Specificity\\ AUC-ROC} & \makecell{Accuracy=0.93 \\Sensitivity=0.90 \\Specificity=0.94 \\AUC-ROC=0.97} & \\	
		\hline
	\end{tabular}}
\end{table*}

\begin{table*}[b]
	\caption{Overview of applications of classical features-based methods in sonographic diagnosis of thyroid cancer}
	\centering
	\label{table4}
	\resizebox{\textwidth}{!}{\begin{tabular}{| c | c |c | c | c|c|c|c|}
		\hline
		\textbf{Citation} & \textbf{Study Purpose} & \multicolumn{3}{c|}{\textbf{Dataset}} &\textbf{ Method} & \textbf{Measures }& \textbf{Results or main findings}\\
		& &\textbf{General information} & \textbf{Number of patients} & \textbf{2D or 3D} & & & \\
		\hline
		[14]-2019 & Thyroid segmentation & \makecell{1.Local (675 US slices) \\ 2.Opencas (1600)} &\makecell{1. 6 healthy \\2. 16 healthy} &\makecell{1. 2D \\ 2. 2D} & \makecell{Wavelet texture \\ +kmeans algorithm} & Dice coefficient &\makecell{1.DC:89.66\% \\ 2.DC:86.89\%} \\
		\hline		
		[19]-2019 & \makecell{Nodules classification} & Local (60 thyroid nodules) & 60 & 2D & \makecell{ Two-Threshold Binary Decomposition\\ + Random Forests (RF)\\ and Support Vector Machine (SVM) classifiers} & \makecell{Accuracy\\ Sensitivity\\ Specificity\\ AUC-ROC} & \makecell{1. Acc=95\%, Sensitivity=95\%, \\ Specificity=95\%, AUC-ROC=0.971 for RF \\2. Acc=91.6\%, Sensitivity=95\%, \\ Specificity=90\%, AUC-ROC=0.965 for SVM} \\
		\hline
		[11]-2019 & \makecell{Automated malignancy detection} & Local (2032) & 4765 & 2D & \makecell{Clinical features: size, echogenicity,\\ margins, internal composition, \\shape, aspect ratio, vascularity, capsule, \\hypoechoic halo, cervical lymph nodes, \\calcifications, real-time elastography grade\\Nine classifiers (Random forest, support vector machines,\\ AdaBoost, neural network, k-nearest neighborhood, \\naive Bayesian, convolutional neural network} & \makecell{AUC\\ Sensitivity\\ Specificity\\ Accuracy} & \makecell{AUC-ROC=0.938\\Sensitivity=89.1\%\\Specificity=85.3\%\\Acc=85.7\%} \\		
		\hline
		[15]-2018 & \makecell{Thyroid, lumen and external \\elastic laminae segmentation} & \makecell{1. Opencas\\2. 2011 MICCAI workshop} & \makecell{1.16\\2.10} & \makecell{1.3D\\2.3D} & \makecell{A combination of iterative random walks solver,\\ random forest learning model\\ and gradient vector flow based\\ interframe belief propagation technique} & \makecell{Jaccard index\\ Percentage of Area Difference\\ Hausdorff Distance} &  \makecell{1.Jaccard index= 0.908\\2.Jaccard index=0.937}\\
		\hline
		[27]-2018 & Thyroid nodule detection & \makecell{1.Local\\2.88 images from DDTI dataset} & \makecell{1.50\\2.Not mentioned} & \makecell{1.2D\\2.2D} & \makecell{Speckle reduction using Neutrosophic \\Nakagami Total Variation,\\ small rectangular region extraction and \\ nodule segmentation using\\ Neutrosophic Distance Regularized Level Set} & \makecell{True Positive\\ False Positive\\Dice Coefficient\\Overlap Metric\\Hausdroff Distance}& \makecell{True Positive=95.92\%\\ False Positive=7.04\%\\Dice Coefficient=93.88\%\\Overlap Metric=91.18\%\\Hausdroff Distance=0.52} \\
		\hline
		[12]-2017 & Thyroid cancer detection & Local(203 nodules) & 187 & 2D & \makecell{Sonographic features+ \\ReliefF feature selection+ \\classification methods: extreme \\learning machine (ELM), SVM \\and bp neural network} &\makecell{Accuracy\\AUC-ROC\\Sensitivity\\Specificity}& \makecell{Accuracy=87.72\%\\AUC-ROC=0.86\\Sensitivity=78.89\%\\Specificity=94.55\%} \\		
		\hline
		[20]-2017 & Thyroid cancer detection& Local(610 images) & 543 & 2D & \makecell{Region-based active contours+\\ morphological and texture features+\\Classifiers: ANN and SVM} & \makecell{sensitivity\\specificity\\positive predictive value\\negative predictive value\\Youden index\\accuracy}& \makecell{sensitivity=100.00\%\\specificity=87.88\%\\positive predictive value=80.95\%\\negative predictive value=100.00\%\\Youden index=87.88\%\\accuracy=92.00\%} \\
		\hline
		[13]-2017 & Thyroid cancer detection & Local(308)& Not mentioned & 2D & \makecell{Sonographic features: texture, contour definition, \\Hypoechogenic halo, Nodule Echogenicity, \\Internal composition, Sphericity, Calcifications\\classifiers: Random Forest, Support Vector Machine \\and Logistic Regression}& \makecell{AUC-ROC\\sensitivity\\specificity}& \makecell{AUC-ROC=0.80\\sensitivity=83\%\\specificity=77\%} \\	
		\hline
		[24]-2016 & Thyroid cancer detection & \makecell{Local (242 \\ thyroid HRUS images)} & 223 & 2D & \makecell{1. Image pre-processing using CLAHE \\2. Extracting Fuzzy, Shannon, Renyi, HOS,\\ Kapur and Vajda entropy features \\from Gabor transform coefficients\\3. Feature space reduction and data normalization using\\ locality sensitive discriminant analysis (LSDA)\\ and ChiMerge discretization\\4. Feature ranking using Relief-F\\5. Classification using SVM, KNN and MLP} & Accuracy& Accuracy=94.3\% \\
		\hline
		[25]-2016 & Thyroid cancer detection & Local (118 US images) & 59 & 2D & \makecell{ Feature set: histogram, intensity differences,\\ elliptical fit, gray-level co-occurrence matrices\\ and gray-level run-length matrices\\3. Classifier: SVM} & \makecell{Accuracy\\AUC-ROC}& \makecell{Accuracy=98.3\%\\AUC-ROC=0.986} \\
		\hline
		[16]-2016 & \makecell{Thyroid nodules  segmentation} & \makecell{Local(B-mode thyroid US\\ image of 42 subjects)} & 42 & 2D & \makecell{Spatial Neutrosophic Distance Regularized \\Level Set method based on\\ Neutrosophic L-Means clustering} & \makecell{true positive rate\\ false positive rate\\ mean absolute distance\\ Hausdorff distance\\Dice metric}& \makecell{true positive rate= 95.45\%\\ false positive rate= 7.32\%\\ mean absolute distance= 1.8\\ Hausdorff distance= 0.7\\Dice metric=94.25\%}\\
		\hline
		[17]-2015 & \makecell{Multi-organ segmentation of thyroid, \\carotid artery, muscles and trachea} & \makecell{1. Local (34 US images in B-mode)\\2. Local (18 US images in B-mode\\from another hospital} & \makecell{1.12\\2.Not mentioned } & \makecell{1.2D\\2.2D} & \makecell{Using speckle-related pixels\\ to quantize the input US image} & Dice coefficient& \makecell{1.DSC for thyroid, carotid, muscles and trachea: \\0.85, 0.87, 0.84, 0.86\\ 2.DSC for thyroid, carotid, muscles and trachea: \\0.81, 0.88, 0.76, 0.87 } \\     
		\hline
		[56]-2013& Thyroid nodule detection & 40 & 40 & 2D & \makecell{Mean intensity has been used \\as the texture feature-\\ first-order statistical features\\ evaluation using t-test and z-test}& \makecell{Accuracy\\ Sensitivity \\ Specificity}& \makecell{Acc=83\%\\Sens=90\%\\Spec=75\%} \\
		\hline
		[23]-2013& \makecell{Thyroid cancer detection \\based on Elastography} & Local(125 color elastograms) & 125 & Elastograms & \makecell{1. Lesion region extraction by radiologist\\2. Statistical and texture features extraction\\3.Feature selection and SVM classification} & \makecell{Sensitivity\\Specificity\\Accuracy}& \makecell{Sensitivity=94.6\%\\Specificity=92.8\%\\Accuracy=93.6\%} \\
		\hline
		[21]-2012& \makecell{Thyroid cancer detection} & \makecell{Local (400 benign and \\400 malignant cases)} & 20 & 3D & \makecell{A combination of DWT (Discrete Wavelet Transform) \\ and texture based features (gray level \\ co-occurrence matrix) + AdaBoost classifier} & \makecell{TN, FN, TP, FP, \\ Acc, Sensitivity, \\ Specificity and AUC} & \makecell{AUC-ROC= 1\\ Acc, sensitivity and specificity= 100\%} \\
		\hline
		[26]-2010 & \makecell{Thyroid nodule classification} & Local (1639 ROI) & 61 & 2D & \makecell{Extracting 78 texture features from ROIs-\\ Train 6 SVMs for 6 types of nodules} & Accuracy& 100\% \\
		\hline
		[18]-2010& \makecell{Thyroid segmentation}& Local (20 US images) & 20 & 2D & \makecell{1. Locating Probable Thyroid Region\\2. Adaptive Weighted Median Filter\\3.Remove the redundancy by\\ Morphological Operation\\4.Feature extraction using Haar Wavelet,\\Histogram feature, BDIP Feature and NMSID Feature\\5.Training RBF neural network to \\classify each block into \\thyroid gland and non-thyroid gland}& \makecell{Accuracy\\Sensitivity\\Specificity\\PPV\\NPV}& \makecell{Accuracy=96.52\\Sensitivity=91.58\\Specificity=97.61\\PPV=89.14\\NPV=98.04} \\
		\hline
		[22]-2009& \makecell{Malignancy risk evaluation \\ of thyroid nodules} & Local (85)& 85 & 2D & \makecell{A combination of morphological \\ and wavelet features + SVM \\ and Probabilistic neural networks (PNNs)} & \makecell{AUC-ROC, Sensitivity, \\ Specificity and Likelihood ratio} & \makecell{AUC-ROC=0.88 \\Sensitivity=0.93 \\Specificity=0.96 \\Likelihood ratio=23.2}  \\
		\hline
	\end{tabular}}
\end{table*}

\end{document}